\documentclass{ws-p9-75x6-50}

\def\vr{{\bf r}}
\def\vrp{\vr^\prime}

\def\vp{{\bf p}}
\def\vA{{\bf A}}
\def\nvr{{\bf r}_1,\cdots{\bf r}_N}
\def\rp{r^\prime}
\def\Deth{\Delta\theta}

\begin{document}

\title{Variational wave functions of a vortex in cyclotron motion}

\author{Jian-Ming Tang}

\address{Department of Physics, University of Washington, \\ P. O. Box 351560, Seattle, WA 98195-1560, USA\\E-mail: jmtang@u.washington.edu}

\maketitle

\abstracts{
In two dimensions the microscopic theory, which provides a basis for
the naive analogy between a quantized vortex in a superfluid and an
electron in an uniform magnetic field, is presented. A one-to-one
correspondence between the rotational states of a vortex in a cylinder
and the cyclotron states of an electron in the central gauge is
found. Like the Landau levels of an electron, the energy levels
of a vortex are highly degenerate. However, the gap between two
adjacent energy levels does not only depend on the quantized
circulation, but also increases with the energy, and scales with the
size of the vortex.  }

It has become common wisdom to treat a quantized vortex in a
two-dimensional superfluid like an electron in an uniform magnetic
field.\cite{Hatsuda94} This analogy is based on the fact that the
dynamics of a vortex in an ideal classical fluid is determined by the
Magnus force and the effective mass of the vortex.\cite{Lamb32} Like
the Lorentz force, the Magnus force is a transverse force whose
magnitude is proportional to the velocity of the vortex relative to
the background fluid. The effective mass is related to the fluid
displaced by the motion of the vortex, and is sensitive to the
structure of the vortex core. The trajectory of a classical vortex
therefore consists of a circular orbit around a guiding
center. However, there is no quantum theory starting from first
principles which supports the existence of such cyclotron motion for a
quantized vortex in a superfluid. Experimental evidence for these
collective modes has been shown in a more complicated situation,
where the cyclotron motion is coupled to the standing wave along the
vortex filament.\cite{Ashton79,Karrai92-2}

It is the purpose of the present study to provide a microscopic theory
for this phenomenological analogy, and show that the rotational states
of a quantized vortex form highly degenerate energy levels similar to
the Landau levels in the integer quantum Hall effect. In order to
describe these rotational states, which are collective modes in a
many-body system, I construct variational wave functions. It was
suggested by Hill and Wheeler\cite{Hill53} and also by Redlich and
Wigner\cite{Redlich54} that the wave function of a collective
excitation in a nucleus could be constructed by integrating out the
surface variables, which specify the position and the orientation of
the nucleus. A quasi-harmonic oscillator wave function was proposed
for the weighting function of the integration. Peierls, Yoccoz and
Thouless later pointed out that the weighting function could be
determined from the variational principle.\cite{Peierls57,Peierls62}
Applying their scheme on a vortex, I calculate the energy spectrum and
the weighting functions, which can be thought as the single-particle
``wave functions'' of a vortex, in a finite system of dilute
bosons. The results are found to be consistent with the analogy, except
that the energy gap between two neighboring levels is not a constant,
and scales logarithmically with the size of the vortex.

I start with a system of $N$ bosons in a disk geometry with radius
$R$. The Hamiltonian contains interactions and a confining potential
representing the boundary of the system. A static vortex, with a
quantized circulation $h/m$, located at a certain point $\vr_0$ in
space can be described by a Feynman wave function,\cite{Feynman55}
\begin{equation}
\Psi_{\vr_0}=\left[\prod_{j=1}^Ng(|\vr_j-\vr_0|)e^{i\phi(\vr_j,\vr_0)}\right]\Psi_{\rm gs}(\nvr) \;,
\end{equation}
where $g(r)$ is a cut-off function to regularize the kinetic energy
near the core, and $\phi(\vr,\vr_0)$ is the velocity potential. An
image vortex with an opposite circulation is present at
$(R/r_0)^2\vr_0$ so that the normal component of the velocity field
vanishes at the boundary. The wave functions representing the
rotational states are constructed as linear combinations of Feynman
wave functions centered at different positions,
\begin{equation}
\Psi_{n,l}=\int d^2\vr_0 f_{n,l}(r_0)e^{il\theta_0}\Psi_{\vr_0}(\nvr) \;,
\end{equation}
where $n$ is the principal quantum number, $l$ is the angular
momentum, and $f_{n,l}(r_0)$ is the radial part of the weighting
function that needs to be determined. This trial function $\Psi_{n,l}$
possesses the proper symmetry of a quantum state because it does not
contain the vortex position $\vr_0$ explicitly.  The single-particle
aspect of the collective motion is represented by
$f_{n,l}(r_0)e^{il\theta_0}$. Minimizing the energy integral leads to
an integral equation for $f_{n,l}(r_0)$,
\begin{equation}
\int dr_0\left[K_l(\rp_0,r_0)-\varepsilon_{n,l}J_l(\rp_0,r_0)\right]f_{n,l}(r_0)=0 \;, \label{eq:int}
\end{equation}
where $J_l$ is the overlap between two Feynman wave functions
separated by a finite distance, $K_l$ is the same overlap weighted by
the kinetic energy, and $\varepsilon_{n,l}$ is the variational energy
relative to the ground-state energy. The interaction potential is
hidden inside the ground state, and does not show up explicitly in
Eq.~(\ref{eq:int}). In the simplest approximation, in which the
cut-off function and the ground state are both set to unity, a
calculation on the overlap integrals gives
\begin{eqnarray}
J_l(\rp_0,r_0) & = & r_0\int d\Deth_0\, m\, S_l(\rp_0,r_0,\Deth_0) \;, \label{eq:ker1}\\
K_l(\rp_0,r_0) & = & r_0\int d\Deth_0\left(\ln\frac{2R}{d}+\beta\right)S_l(\rp_0,r_0,\Deth_0) \;, \\
S_l(\rp_0,r_0,\Deth_0) & = & \exp\left\{-\!\left(\frac{1}{2}\ln\frac{2R}{d}+\alpha\right)d^2-i\left[r_0\rp_0\sin\Deth_0-(N-l)\Deth_0\right]\right\} , \label{eq:ker2}
\end{eqnarray}
where $\alpha=1.0074$, $\beta=-0.30685$, $d=|\vr_0-\vrp_0|$,
$\Deth_0=\theta_0^\prime-\theta_0$, and $m$ is the mass of the
particle. The unit length is set to be the interatomic spacing
$\sigma\equiv R/\sqrt{N}$, and $\hbar$ is set to unity. The integral
equation, Eq.~(\ref{eq:int}) with
Eqs.~(\ref{eq:ker1})--(\ref{eq:ker2}), is exactly solvable as an
eigenvalue problem for $J_l^{-1}K_l$. It is found numerically that the
variational energy $\varepsilon_{n,l}$ is independent of the angular
momentum $l$. Numerical solutions of $\varepsilon_n$ and
$f_{n,l}(r_0)$ are summarized in Fig.~1 and Fig.~2.

\begin{figure}
\epsfxsize=3.3in
\centerline{\epsfbox{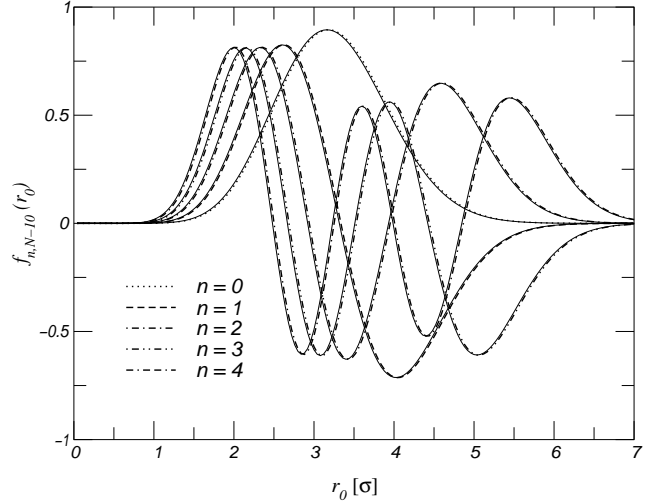}}
\caption{An example of the weighting functions, shown by the solid lines
and normalized as $\int dr_0|f(r_0)|^2=1$, representing the vortex states
with angular momentum $l=N-10$. The dotted lines show the corresponding
radial wave functions of an electron in the central gauge with angular
momentum $l_z=-10$. \label{fig:wf10m10}}
\end{figure}

\begin{figure}
\epsfxsize=3.3in
\centerline{\epsfbox{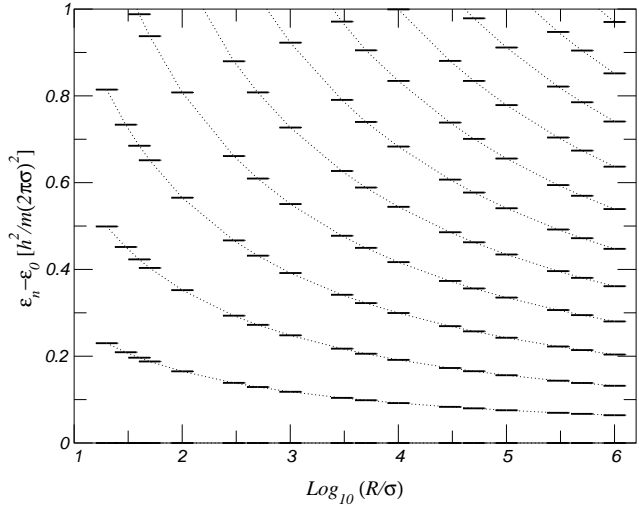}}
\caption{The energy spectrum corresponding to the cyclotron motion
of a vortex relative to the lowest energy level. The dotted lines show
the scaling of each energy level with the system radius $R$ ranging
from $20$ to $10^6$. The lowest energy level approximately scales as
$\ln(R/\sigma)$ corresponding to the formation energy of a static
vortex.
\label{fig:engsp}}
\end{figure}

The motion of a vortex can now be understood by comparing the
weighting functions in Fig.~\ref{fig:wf10m10} to the eigenfunctions of
the Hamiltonian of an electron moving in the $x$-$y$ plane with an
uniform magnetic field $B$ pointing in the $z$ direction,
\begin{equation}
H=\frac{1}{2m}(\vp-e\vA)^2=\frac{\vp^2}{2m}+\frac{m}{2}\left(\frac{\omega}{2}\right)^2(x^2+y^2)+\frac{\omega}{2}L_z \;, \label{eq:electron}
\end{equation}
where $\omega=|e|B/m$ is the cyclotron frequency, $L_z$ is the angular
momentum operator perpendicular to the plane, and the central gauge
$\vA=(-yB/2,xB/2)$ is chosen so that $L_z$ commutes with the
Hamiltonian. The radial wave function $\psi_{n,l_z}(r)$ in the $n$-th
Landau level with angular momentum $l_z$ is identical to the weighting
function $f_{n,N+l_z}(r_0)$ up to a proper normalization, if the unit
length is chosen as $\sqrt{2\hbar/m\omega}$. This one-to-one
correspondence immediately tells us that the motion of a vortex is
indeed cyclotron-like, and the degeneracy of each energy level in
Fig.~\ref{fig:engsp} is one per $\pi$ unit area, since the degeneracy
of a Landau level is $eB/h$ per unit area.

The energy spectrum of a vortex, however, is different from the
spectrum of an electron. The Hamiltonian in Eq.~(\ref{eq:electron})
basically describes a particle in a harmonic potential in the rotating
frame, and has a fixed energy scale $\hbar\omega$ set by the magnetic
field. In Fig.~\ref{fig:engsp} the level spacing of a vortex is
increasing with energy, and scales down logarithmically with the size
of the vortex. Both these features are attributed to the long-ranged
phase coherence, which causes the logarithmic size-dependence of the
overlap in Eq.~(\ref{eq:ker2}) between two Feynman wave functions. The
small difference among level spacings represents a breakdown of the
analogy because one cannot simply add an attractive potential in
Eq.~(\ref{eq:electron}) to mimic the spectrum due to the fact that the
degeneracy of the Landau level will be broken by this extra
interaction. The scaling of the level spacing can still be
incorporated into the single-particle picture by requiring that the
effective mass of the vortex scales with the size accordingly.

In conclusion, I have presented a quantum theory to support the
picture that a quantized vortex behaves like an electron in a magnetic
field. Vortex states with different angular momenta form highly
degenerate levels.

\section*{Acknowledgments}
It is a pleasure to thank David Thouless for initiating the idea of
this work and stimulating discussions. The author is also grateful to
Ping Ao for valuable comments and suggestions. This work was partially
supported by NSF DMR-9815932.


\begin{thebibliography}{1}

\bibitem{Hatsuda94}
M. Hatsuda, S. Yahikozawa, P. Ao, and D.~J. Thouless, Phys. Rev. B
{\bf 49}, 15870 (1994).

\bibitem{Lamb32}
S.~H. Lamb, {\em Hydrodynamics} (The University Press, Cambridge,
1932).

\bibitem{Ashton79}
R.~A. Ashton and W.~L. Glaberson, Phys. Rev. Lett. {\bf 42}, 1062
(1979).

\bibitem{Karrai92-2}
K. Karrai et al., Phys. Rev. Lett. {\bf 69}, 355 (1992).

\bibitem{Hill53}
D.~L. Hill and J.~A. Wheeler, Phys. Rev. {\bf 89},  1102  (1953).

\bibitem{Redlich54}
M.~G. Redlich and E.~P. Wigner, Phys. Rev. {\bf 95},  122  (1954).

\bibitem{Peierls57}
R.~E. Peierls and J. Yoccoz, Proc. Roy. Soc. {\bf A 70}, 381 (1957).

\bibitem{Peierls62}
R.~E. Peierls and D.~J. Thouless, Nucl. Phys. {\bf 38}, 154 (1962).

\bibitem{Feynman55}
R.~P. Feynman,  in {\em Progress in Low Temperature Physics},
edited by C.~J. Gorter (Elsevier Science Publishers B.V., Amsterdam, 1955),
Vol.~1, Chap.~2, pp.\ 17--53.

\end{thebibliography}
\end{document}